\documentclass{article}
\usepackage{spconf,amsmath,graphicx,hyperref,dirtytalk}
\usepackage{adjustbox}
\usepackage{booktabs}
\usepackage{float}
\usepackage{algorithm}
\usepackage{algpseudocode}
\usepackage{subfig}
\usepackage{amssymb}
\newcommand{\ra}[1]{\renewcommand{\arraystretch}{#1}}

\ninept
\title{Learning Music Representations with wav2vec 2.0}
\name{Alessandro Ragano$^{1}$, \qquad Emmanouil Benetos$^{2}$, \qquad Andrew Hines$^{1}$ \thanks{This publication has emanated from research conducted with the financial support of Science Foundation Ireland (SFI) under Grant Number 17/RC-PhD/3483 and 17/RC/2289\_P2 and was supported by The Alan Turing Institute under the EPSRC grant EP/N510129/1. EB is supported by a Turing Fellowship.}}

\address{$^{1 }$School of CS, University College Dublin, Ireland \\ $^{2}$School of EECS, Queen Mary University of London, UK}

\begin{document}
%
\maketitle
\begin{abstract}
Learning music representations that are general-purpose offers the flexibility to finetune several downstream tasks using smaller datasets. The wav2vec 2.0 speech representation model showed promising results in many downstream speech tasks, but has been less effective when adapted to music. In this paper, we evaluate whether pre-training wav2vec 2.0 directly on music data can be a better solution instead of finetuning the speech model. We illustrate that when pre-training on music data, the discrete latent representations are able to encode the semantic meaning of musical concepts such as pitch and instrument. 
Our results show that finetuning wav2vec 2.0 pre-trained on music data allows us to achieve promising results on music classification tasks that are competitive with prior work on audio representations. In addition, the results are superior to the pre-trained model on speech embeddings, demonstrating that wav2vec 2.0 pre-trained on music data can be a promising music representation model.
\end{abstract}
\begin{keywords}
music representations, self-supervision, pre-training
\end{keywords}
\section{Introduction}
\label{sec:intro}
Learning feature representations with deep architectures has shown remarkable success over hand-crafted features in Music Information Retrieval (MIR)~\cite{humphrey2013feature}. Approaches such as transfer learning from music auto-tagging~\cite{choi2017transfer,van2014transfer,pons2019musicnn} allows someone to pre-train neural networks using large datasets and extracting features for downstream MIR tasks such as instrument classification or genre recognition. In this way, downstream MIR tasks can be solved using smaller annotated datasets, which is desired since labeling is costly and difficult to achieve. One issue with
auto-tagging models is that they require very large annotated datasets that are still difficult to obtain. 
To overcome the need for large annotated datasets, new music representation techniques have emerged that do not directly use waveform-related labels emerged. For example, pre-training from language models~\cite{castellon2021codified} or using noisy language descriptors of the musical content~\cite{manco2022learning}.
 
A different approach that is based on using proxy tasks to learn representations is self-supervised learning (SSL), where information from input data is extracted to provide labels. This is advantageous since labels can be generated automatically without requiring human intervention. Some SSL models have been proposed for music representations showing competitive performance in several downstream MIR tasks~\cite{wu2021multi,zhu2021musicbert,carr2021self,spijkervet2021contrastive}. Beyond music representation learning, SSL models have grown popularity for speech representations and downstream tasks such as speaker identification, automatic speech recognition, phoneme recognition, and speech translation~\cite{mohamed2022self}. Examples of speech SSL models include wav2vec 2.0~\cite{baevski2020wav2vec} which is a contrastive learning-based approach where the model learns to distinguish a target sample (positive) from distractors (negative). The original model was pre-trained on the LibriSpeech dataset~\cite{panayotov2015librispeech} and its success is highlighted by the ability to retain high performance even when dedicated datasets for downstream tasks are very small, e.g., 10 minutes only for speech recognition~\cite{baevski2020wav2vec} or 1000 observations for non-intrusive speech quality assessment~\cite{becerra2022exploring}.

The wav2vec 2.0 SSL model has been extensively evaluated for speech tasks. However, its adaptation to music tasks (such as pitch classification or instrument classification) has been explored so far without success, as shown in two studies. In the NeurIPS challenge HEAR~\cite{turian2022hear}, wav2vec 2.0 embeddings are extracted from the model pre-trained on the LibriSpeech dataset and are used as input features without finetuning. Their performance is relatively low in music tasks, even if wav2vec 2.0 speech embeddings can still represent some musical concepts to some degree, such as pitch~\cite{turian2022hear}. Wang et al.~\cite{wang2022towards} have also evaluated wav2vec 2.0 outside of the speech domain. In this case, the authors found that wav2vec 2.0 did not perform well when pre-trained on AudioSet~\cite{gemmeke2017audio}, possibly due to the limitation of the masked prediction objective of learning from a dataset more complex than LibriSpeech~\cite{wang2022towards}.

An approach that is still unexplored is pre-training wav2vec 2.0 on music data only. The transferability of deep networks becomes more challenging when the source and the target tasks have different domains~\cite{yosinski2014transferable} and it has been shown that wav2vec 2.0 might be sensitive to a domain shift. For example, pre-training wav2vec 2.0 with cross-lingual datasets improves performance of ASR systems~\cite{conneau2020unsupervised} and finetuning with non-English languages shows a performance drop for speech quality assessment~\cite{becerra2022exploring}. 

In this paper, we study whether the domain shift between the pre-trained model and the downstream tasks observed can cause this performance drop in music tasks as reported in the studies above. We explore further the capacity of wav2vec 2.0 features in non-speech tasks asking the following questions: 
\begin{enumerate}
    \item Does wav2vec 2.0 pre-trained on music encode meaningful music representations, i.e. related to musical concepts such as pitch or instruments?
    \item Is it possible to obtain competitive performance on MIR tasks when finetuning wav2vec 2.0 pre-trained on music? 
    \item Can we establish if wav2vec 2.0 is a potential candidate model for music tasks other than speech?
\end{enumerate}

The paper is structured as follows. In Section 2 we illustrate how we pre-train wav2vec 2.0 with music data. 
Section 3 is dedicated to the analysis of the features learned by wav2vec 2.0. We show whether the information encoded in the codebooks is related to music labels and we compare the encoded representations in the continuous layers of wav2vec 2.0 pre-trained on music with the information encoded in the original speech model.
Section 4 shows the results of finetuning wav2vec 2.0 pre-trained on music on two MIR tasks: instrument classification and pitch classification\footnote{\label{note1}In this paper, we use the term pitch classification since the NSynth dataset is made of isolated note segments. This is different from the more common term "pitch detection" where note segments are not isolated.}. In Section 5 we discuss whether wav2vec 2.0 pre-trained on music provides promising potential for broader downstream MIR tasks.

\section{Method}
\subsection{Pre-Trained Model}
The wav2vec 2.0 model can be summarized in the following blocks:
\begin{enumerate}
    \item A \textit{feature encoder} $f:\mathbb{X}\mapsto \mathbb{Z}$ that converts input audio chunks of 20 milliseconds $X$ into a sequence of latent speech representations $\mathbb{Z}=\{z_1, z_2, ..., z_T\}$ for $T$ timesteps. The encoder consists of 7 1D convolutional layers, each with 512 filters.  
    \item A \textit{context network} $g:\mathbb{Z}\mapsto \mathbb{C}$ based on the Transformer architecture~\cite{vaswani2017attention} that builds context representations for each audio segment that capture the entire audio sequence $\mathbb{C}=\{c_1, c_2, ..., c_T\}$ 
    \item A \textit{quantization module} that transforms encoder output representations into discrete speech representations $\mathbb{C}\mapsto \mathbb{Q}$. The discrete latent features are learned with product quantization and are needed to create targets for the loss function, but they are not used as input for the context network. A vector that concatenates an entry from each of the 2 codebooks is linearly transformed to get the quantized representations $\mathbb{Q}=\{q_1, q_2, ..., q_T\}$. The Gumbel-Softmax is used to choose the codebook entries in a differentiable way.
    \item A \textit{contrastive loss function} is used to learn how to identify the true quantized speech representation from 100 quantized negative samples that are uniformly sampled. Given an audio chunk at time step $t$, the model compares the cosine similarity between the Transformer output at time step $t$ and the quantized speech representation in the same step $t$ against the similarity with negative distractors. The high similarity with the negative samples is penalized by contrastive loss. The latent speech representation at the time step $t$ created by the feature encoder is masked before being fed to the Transformer-based context network. Negative samples are sampled from other masked time steps of the same utterance.     
\end{enumerate}
We use the BASE model configuration~\cite{baevski2020wav2vec}, which consists of 12 Transformer blocks and produces $768$-dimensional feature vectors.

To learn music representations, we pre-trained wav2vec 2.0 on the MusicNet dataset~\cite{Thickstun2017}. MusicNet consists of $\approx34$ hours of audio across 330 classical music recordings provided as raw waveforms, covering 11 musical instruments. To pre-train wav2vec 2.0 we use the fairseq toolkit~\cite{ott2019fairseq}. The data is split into overlapped segments of 20 seconds, whose length is recommended in the fairseq repository instructions. To increase the dataset size we take overlapped segments with hop size equal to 10 seconds collecting $\approx65$ hours of audio in total. The dataset that we used to pre-train represents only $\approx7\%$ of the LibriSpeech dataset size which is the one used to pre-train wav2vec 2.0 for speech~\cite{baevski2020wav2vec}. However, we will show that this is sufficient to address whether wav2vec 2.0 learns meaningful music representations. The model was trained for 1790 epochs and it took 7 days on the NVIDIA A100 64GB GPU. 

\subsection{Finetuning}
Evaluation of downstream tasks is performed on the NSynth dataset \cite{engel2017neural} using the original train, validation, and test splits. The NSynth dataset includes 305,979 samples of 4 seconds. Two tasks are evaluated on this dataset, pitch classification~\ref{note1} and instrument classification. Pitch labels on the isolated note recordings are provided as MIDI numbers. Instrument labels represent the instrument family and include the following 11 instruments: bass, brass, flute, guitar, keyboard, mallet, organ, reed, string, synth\_lead, vocal. Notice that synth\_lead is only present in the training set, which makes validation and test splits made of 10 instrument classes. 

The output of the last Transformer block is a matrix of size $(n \times l)$ where $n$ is the number of time frames and $l$ is the size of the feature vector equal to $768$. To remove the time dimension, we simply average across time, obtaining an $l$-dimensional vector at the output.
The latter is connected to a linear layer that consists of the number of output neurons equal to the number of classes of the task: 112 neurons for pitch classification and 11 neurons for instrument classification.

The pre-trained wav2vec 2.0 model with music data is used in 3 different configurations for the downstream tasks: 1) finetuning (FT1) the entire network, 2) finetuning (FT2) the context network (Transformer) while keeping the feature encoder frozen, 3) Freezing both feature and context networks and doing a simple feature extraction (FE) which consists of training only the output linear layer.
Finetuning on models FT1 and FT2 is performed using the Adam optimizer with a learning rate of 0.00001 for the pre-trained part and 0.0001 for the output layer. The FE model is trained using the Adam optimizer with a learning rate of 0.001 and only the weights of the output linear layer are optimized. In all 3 configurations, training is stopped if the average loss in the validation set did not decrease for 10 epochs. The cross-entropy loss is used for classification.

\section{Feature Analysis}
Our first research question in Section 1 asked whether wav2vec 2.0 learns meaningful representations when pre-trained on music data. 
We first explored whether the learned discrete latent representations used in the loss function encode a semantic meaning related to musical concepts.
The discrete representations are an important step in wav2vec 2.0 since learning a finite set of discrete audio units encourages the model not to learn all the variations in the data when minimizing the contrastive loss. We use the NSynth dataset to compute the co-occurrence between both pitch and instrument family labels and the discrete latent features produced by wav2vec 2.0 pre-trained on MusicNet without finetuning.
\begin{figure}[!t]
\centering
\includegraphics[width=0.90\linewidth]{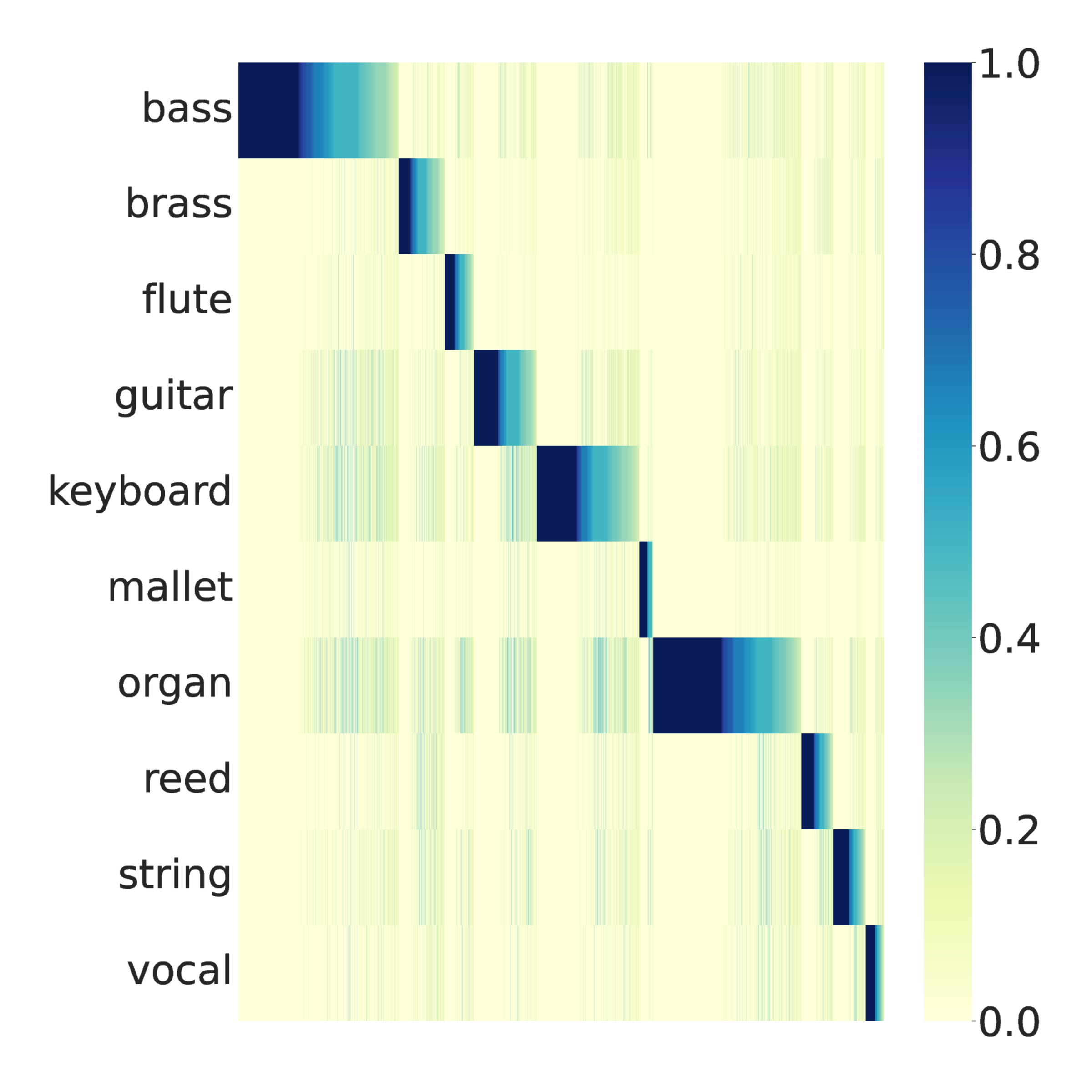}
\caption{Co-occurrence between the discrete latent representations and instrument family labels on the NSynth test set.}
\label{fig:cooc_instr}
\end{figure}

\begin{figure}[!t]
\centering
\includegraphics[width=0.99\linewidth]{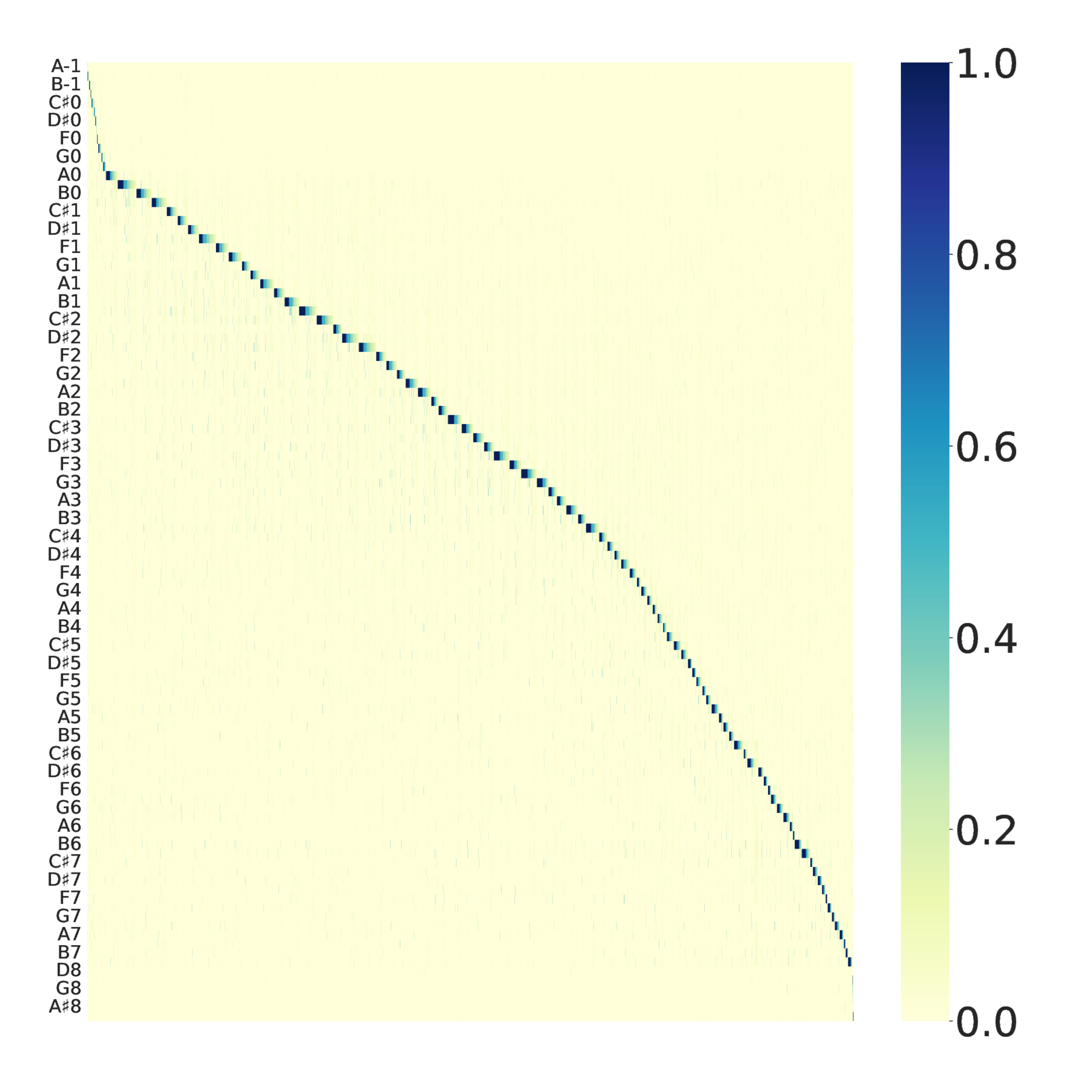}
\caption{Co-occurrence between the discrete latent representations and pitch classes on the NSynth test set.}
\label{fig:cooc_pitch}
\end{figure}

Figure~\ref{fig:cooc_instr} and Figure~\ref{fig:cooc_pitch} show that discrete latent representations specialize in both instrument and pitch classes, respectively. Many latents co-occur with bass, which is the most frequent class in the NSynth test. The discrete latent representations share a similar pattern as the wav2vec 2.0 speech model, where the encoded semantic meaning of the codebooks has been shown to be represented by phonemes~\cite{baevski2020wav2vec}. 

A deeper insight in the analysis of the wav2vec 2.0 features can be obtained by analyzing the Transformer layers. 
Given a masked latent representation $z_t$, the objective of the model is to learn a context representation $c_t$ in order to correctly guess the quantized representation $q_t$ among the negative samples. For this reason, it should be expected that the final layers of the Transformer should have higher similarity with the Transformer input. This behaviour should be observed regardless of the input signal type (speech or music).
To confirm whether the Transformer layers evolve in the pre-trained model as expected, we follow the same approach of Pasad et al.~\cite{pasad2021layer} where they observed this phenomenon occurring in the wav2vec 2.0 pre-trained on speech. We computed the canonical correlation analysis (CCA) between each Transformer layer and the output of the feature encoder. Given a matrix $W \in \mathbb{R}^{n\times k}$ and $Y \in \mathbb{R}^{n\times j}$ with $k<j$, CCA finds two basis such that when the matrices are projected onto the basis their correlation is the highest. More specifically, the CCA is calculated as follows: 
\begin{equation}
\rho_i=\max_{u^{i}_{w}, u^{i}_{y}} corr (Wu^{i}_{w},Yu^{i}_{y}),
\end{equation}

\begin{equation}
CCA(W,Y) = \frac{\sum\limits_{i=1}^{k}\rho_i}{k}
\end{equation}

where $\rho_i$ represents the $i$-th canonical correlation coefficient, $u^{i}_{w}$ and $u^{i}_{y}$ are the vectors found by CCA that maximize the canonical weights, and the final CCA is obtained with the average. In our analysis, the matrices are represented by the feature vectors at each timestep. 
We use a variant projection weighted canonical correlation analysis (PWCCA)~\cite{morcos2018insights} that is less sensitive to perturbation since it uses a weighted mean to assign a higher weight to the correlation coefficients that have more importance.

\begin{figure}[!b]
\centering
\includegraphics[width=0.90\linewidth]{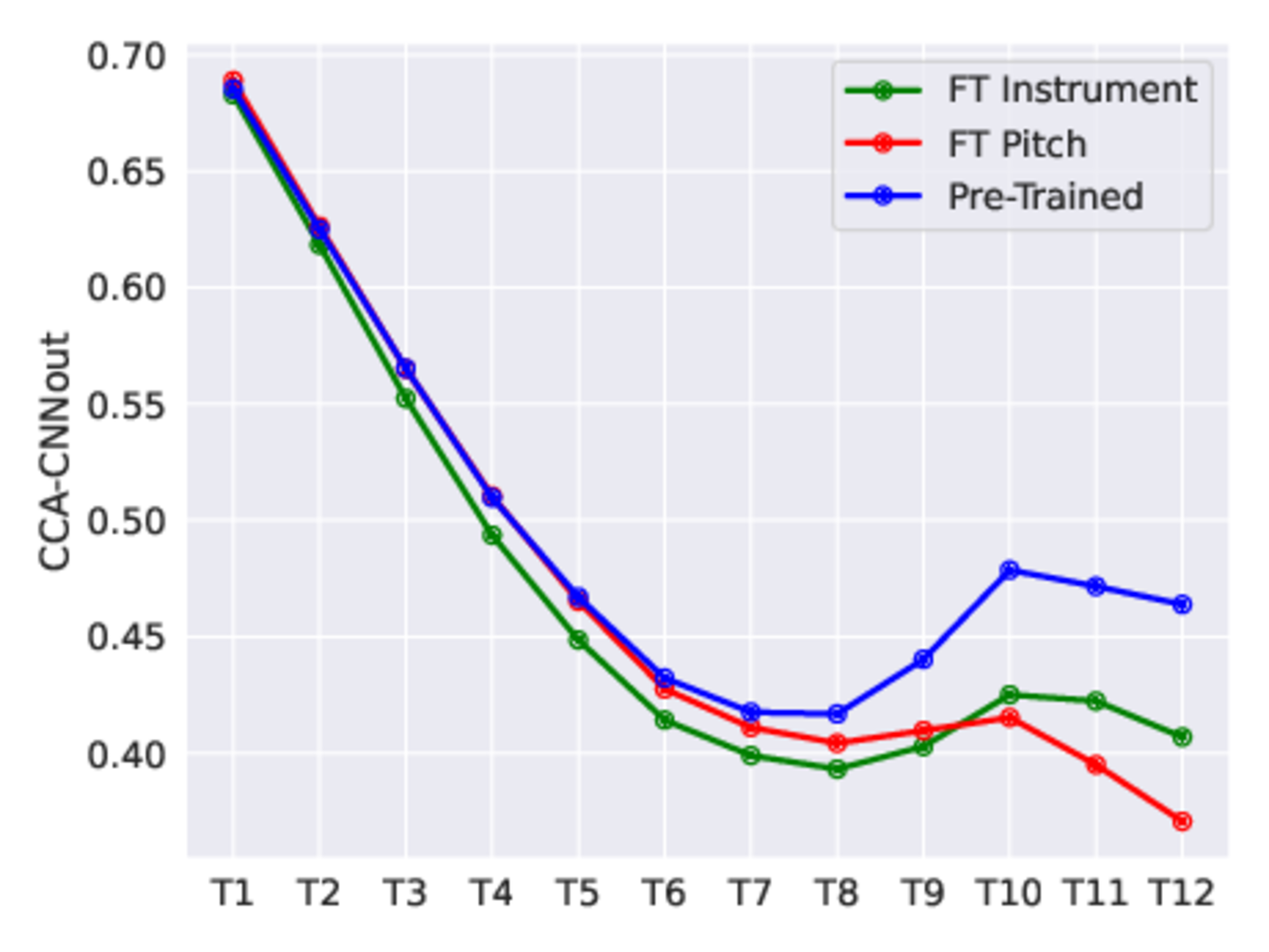}
\caption{Evolution of the Transformer layers of the pre-trained model and the finetuned models using PWCCA between each layer and the output of the feature encoder (CNN).}
\label{fig:cca_cnnout}
\end{figure}

The PWCCA is calculated using the FT1 approach where the feature encoder is frozen, and by using frames extracted from the MusicNet dataset. Due to the high computational effort, we take 4 seconds in the middle of each MusicNet observation using half of the dataset size. 
Figure~\ref{fig:cca_cnnout} shows that the pre-trained model attempts to reconstruct the input features (i.e. the output of the feature encoder) while the similarity between the final layers of the finetuned models tend to be lower than the pre-trained model. This confirms that the evolution of the Transformer layers with respect to the feature encoder output is the same in both the speech model (Pasad et al.~\cite{pasad2021layer}) and the music model (this study), which is aligned with the objective of wav2vec 2.0.

\section{Downstream Tasks}
The second research question in Section 1 asked whether finetuning wav2vec 2.0 pre-trained on music data shows competitive performance in downstream MIR tasks. The performance of wav2vec 2.0 pre-trained on music data is evaluated on pitch and instrument classification using the NSynth test set. The model is compared with  previous works as shown in Table~\ref{tab:w2v_res}. 

For pitch classification we use CREPE~\cite{kim2018crepe} which is the best pitch classifier in the HEAR challenge~\cite{turian2022hear}, SF NFNet-50 which is the best model in a comparison of audio representations reported by Wang et al.~\cite{wang2022towards}, and features extracted from wav2vec 2.0 pre-trained on LibriSpeech which is fundamental to understanding the differences with wav2vec 2.0 pre-trained on music.
All the models, except for CREPE, are designed to learn general purpose audio representations.

For instrument classification, we consider MuLaP~\cite{manco2022learning} which learns music representations by using weak supervision from noisy language descriptors of the musical content, the work of Favory et al.~\cite{favory2021learning} that we call contextual tag embeddings (CTE) where the learned audio representations are aligned to music tags, SF NFNet-50 which is also the best model for instrument classification in the same study from Wang et al.~\cite{wang2022towards}, and the feature extracted from wav2vec 2.0 pre-trained on LibriSpeech.

It should be noted that the models reported in Table~\ref{tab:w2v_res} have some differences that do not allow for direct comparisons such as pre-training datasets, supervision strategies, hyper-parameters, and strategies to use the learned features. However, the choice of the above prior work models helps us to contextualize the results obtained with wav2vec 2.0 pre-trained on music.

The results in Table~\ref{tab:w2v_res} show that pre-training wav2vec 2.0 on music shows comparable results with prior work and improvement over wav2vec 2.0 pre-trained on speech. For pitch classification, finetuning the entire wav2vec 2.0 pre-trained on music achieves the best results together with CREPE and it is the best model among the ones trained to learn general-purpose audio representations. Also, extracting features from the music model shows an 11\% increase over the speech embeddings, indicating that pre-training wav2vec 2.0 on music is the contributing factor to the observed performance improvement. For instrument classification, wav2vec 2.0 pre-trained on music is the second best model after SF NFNet-50 and it shows better results than MuLaP and CTE. We also observe that feature extraction of wav2vec 2.0 from the music model has significant improvement over the speech model, which confirms the positive contribution of the music data used in the pre-training phase.
An important aspect to consider is that we pre-trained wav2vec 2.0 on a relatively smaller dataset which still shows competitive results with the other approaches that are pre-trained on larger datasets.

\begin{table}[!t]
\caption{Performance evaluation using accuracy (\%) on NSynth test set. The results of the baseline model are taken from \cite{turian2022hear}\textsuperscript{+},\cite{wang2022towards}\textsuperscript{++}\cite{manco2022learning}\textsuperscript{+++}}
\centering
\small
\ra{1.0}
\begin{adjustbox}{max width=0.50\textwidth}
\begin{tabular}{@{}llccclll@{}}\toprule

&   Pitch  & Instrument & Pre-training Data \\ \midrule
wav2vec 2.0 Music FE                                                                & 76.0 &   64.0 & Music, $\approx65$ hours \\
wav2vec 2.0 Music FT1                                                               & 82.0 &   70.0 & Music, $\approx65$ hours \\
wav2vec 2.0 Music FT2                                                               & 90.0 &   75.0 & Music, $\approx65$ hours \\
\hline
CREPE~\cite{kim2018crepe}\textsuperscript{+}                                        & 90.0 &         &  \\
SF NFNet-50~\cite{wang2022towards}\textsuperscript{++}                              & 88.0 &   78.2  & Audio, $\approx5800$ hours \\
wav2vec 2.0 Speech FE~\cite{baevski2020wav2vec}\textsuperscript{+}                  & 65.0 &         & Speech,$\approx960$ hours  \\
wav2vec 2.0 Speech FE~\cite{baevski2020wav2vec}\textsuperscript{++}                 & 35.0 &   40.2  & Speech,$\approx960$ hours  \\
MuLaP~\cite{manco2022learning}\textsuperscript{+++}                                 &       &  71.7  &  \\
CTE~\cite{favory2021learning}\textsuperscript{+++}                                  &       &  70.0  & Music, $\approx562$ hours  \\
\bottomrule
\label{tab:w2v_res}
\end{tabular}
\end{adjustbox}
\end{table}

\section{Discussion \& Conclusions}
In this paper, our aim was to study the potential of wav2vec 2.0 on learning meaningful representations from music data.
We pre-trained wav2vec 2.0 on music data and evaluated the model on pitch and instrument classification. We demonstrated that wav2vec 2.0 encodes semantic meaning related to musical concepts in the discrete latent representations and that the Transformer layer behaviour is the same of the speech model. We showed that finetuning wav2vec 2.0 pre-trained on music has significant improvement over the original model pre-trained on speech and other audio-representations models. We posed the question: is wav2vec 2.0 pre-trained on music a potential model for learning general-purpose music representations? Our results and analysis support further application of wav2vec 2.0 with music pre-training for broader downstream MIR tasks. 

Specifically, we propose to extend these findings by performing a direct comparison with the other models and addressing the following: (i) we pre-trained the model using a small dataset which was sufficient for the questions addressed in this paper but not for general-purpose audio representations that require pre-training with much larger datasets;  (ii) the evaluation of the MIR task was conducted using a monophonic dataset (NSynth) so the generalization for downstream polyphonic tasks should be explored; (iii) a broader mix of genres in the pre-trained dataset should be explored as MusicNet includes Western classical music and also non-Western music; (iv) the model hyperparameters were not adjusted or optimized and may be better suited to speech than to music; (iv) the model has been trained for a fixed number of epochs due to GPU capacity constraints but more training epochs can be used e.g. by monitoring the contrastive loss with a validation set.

\bibliographystyle{IEEEbib}
\bibliography{strings,refs}
\end{document}